\documentclass[prl,twocolumn,superscriptaddress,showpacs]{revtex4}

\usepackage{graphicx}
\usepackage{dcolumn}
\usepackage{bm}

\begin{document}

\preprint{ATB-1}

\title{Magnetic excitations in La$_{1.5}$Sr$_{0.5}$CoO$_4$}

\author{L. M. Helme}
\email{l.helme1@physics.ox.ac.uk} \affiliation{ Department of
Physics, Oxford University, Oxford, OX1 3PU, United Kingdom }
\author{A. T. Boothroyd}\affiliation{
Department of Physics, Oxford University, Oxford, OX1 3PU, United
Kingdom }
\author{D. Prabhakaran}\affiliation{
Department of Physics, Oxford University, Oxford, OX1 3PU, United
Kingdom }
\author{F. R. Wondre}\affiliation{
Department of Physics, Oxford University, Oxford, OX1 3PU, United
Kingdom }
\author{C. D. Frost}\affiliation{ISIS Facitlity,
Rutherford Appleton Laboratory, Didcot, U.K.}
\author{J. Kulda}\affiliation{
Institut Laue-Langevin, BP 156, 38042 Grenoble Cedex 9, France }


\begin{abstract}
We report magnetization and neutron scattering measurements of the
half-doped compound La$_{1.5}$Sr$_{0.5}$CoO$_4$, which exhibits a
checkerboard pattern of charge ordering below $\sim$ 800 K. In the
antiferromagnetically-ordered phase below $\sim$ 40 K the spins
are found to be canted in the {\it ab} plane. The spin excitation
spectrum includes spin-wave excitations with a maximum energy of
16 meV, and diffuse magnetic modes at energies around 30 meV.
\end{abstract}

\maketitle


Layered transition metal oxides have been much studied in an
attempt to understand the mechanism of high-T$_c$
superconductivity in the cuprates. Recently there has been
speculation that charge stripe correlations may play an important
role, leading to investigations of compounds such as
La$_{2-x}$Sr$_x$NiO$_4$ which exhibits well defined spin and
charge stripe ordering \cite{tranquada-Nature-1995}.

Here we consider the structurally identical but less-studied
family, La$_{2-x}$Sr$_x$CoO$_4$. Previous studies
\cite{zaliznyak-2000,zaliznyak-2001} of the half-doped member
La$_{1.5}$Sr$_{0.5}$CoO$_4$ have concluded that below $T_c\approx$
800 K the holes introduced by Sr doping form a charge ordered
phase, with a loosely correlated checkerboard arrangement of
Co$^{3+}$ and Co$^{2+}$ ions. It has also been reported that very
slightly incommensurate long range magnetic order occurs below
$\sim$ 60 K \cite{moritomo-1997,zaliznyak-2000}, with a
spin-freezing transition at $T_s\approx$ 30 K
\cite{zaliznyak-2000}.


We report here on polarized- and unpolarized-neutron scattering
measurements and magnetometry studies of single crystals of
La$_{1.5}$Sr$_{0.5}$CoO$_4$. The crystals were grown by the
floating-zone method in Oxford.

Magnetization data were collected using a superconducting quantum
interference device (SQUID) magnetometer with the field parallel
to the {\it ab} plane. DC field-cooled (FC) and zero-field-cooled
(ZFC) data were recorded in a measuring field of 100 Oe.

\begin{figure}[htb]
\includegraphics*{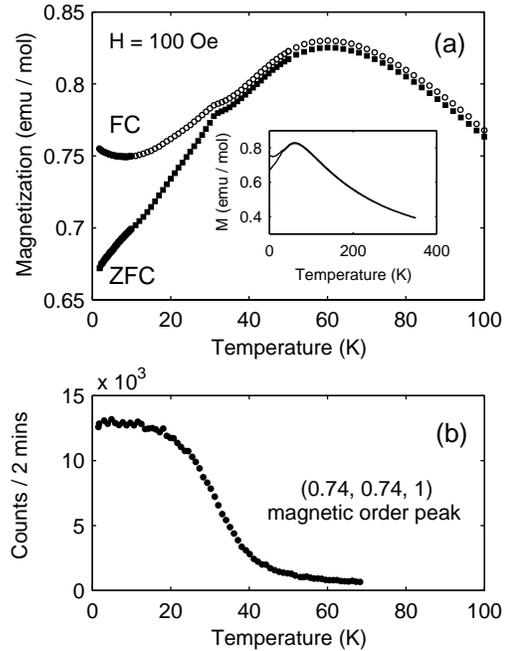}|
\caption{\small (a) Temperature variation of FC and ZFC in-plane
magnetization
(b) Temperature
variation of the amplitude of the magnetic peak at (0.74, 0.74,
1).} \label{fig:squid}
\end{figure}

Figure \ref{fig:squid}a shows the temperature variation of the
in-plane FC and ZFC magnetizations, with the inset showing data up
to 350 K. Our results agree with those of Moritomo et al.
\cite{moritomo-1997}, with a broad maximum in the magnetization
centred at about 60 K. However, we also see a definite splitting
between the ZFC and FC data at low temperatures indicating a
glassy ground-state, and an abrupt change in slope at 30 K. It is
possible that this feature corresponds to a spin-reorientation
transition as observed in La$_{1.5}$Sr$_{0.5}$NiO$_4$
\cite{freeman}.

To examine the magnetic order further we performed
polarized-neutron diffraction on the triple-axis-spectrometer IN20
at the Institut-Laue-Langevin. We studied two rod-shaped crystals
of La$_{1.5}$Sr$_{0.5}$CoO$_4$ (taken from one original crystal)
with a combined mass of 12 g.

Figure \ref{fig:squid}b shows the temperature variation of the
amplitude of a magnetic Bragg peak. This data confirms that
La$_{1.5}$Sr$_{0.5}$CoO$_4$ is magnetically ordered below $\sim$
40 K, but reveals a slow build-up of magnetic correlations with
decreasing temperature below $\sim$ 60 K. It is likely that this
gradual ordering explains the broad maximum in the magnetization
data (fig. \ref{fig:squid}a).

Polarization analysis performed on the magnetic order peaks
revealed that below 30 K the spins lie in the {\it ab} plane at an
angle of $\sim 12^{\circ}$ to the Co-O bonds. We also found
tentative evidence that a small reorientation ($\sim 5 ^{\circ}$)
occurs above 30 K, which would correspond to the kink in fig.
\ref{fig:squid}a. Such a spin reorientation should cause an
anomaly in the temperature variation of the magnetic peak shown in
fig. \ref{fig:squid}b but no anomaly is evident. This may be
because of the small magnitude of the reorientation, or it may
imply a different origin for the magnetization kink.

We studied the spin excitation spectrum both with polarized- and
unpolarized-neutron scattering. The latter was performed with the
MAPS spectrometer at ISIS on a crystal of mass 35.5 g. The spin
correlations were found to be highly two-dimensional (2D), showing
no measureable inter-plane correlations for energies above 4 meV.
For simplicity, therefore, we describe the scattering with respect
to the 2D reciprocal lattice of the square CoO$_2$ planes.

\begin{figure}
\includegraphics*{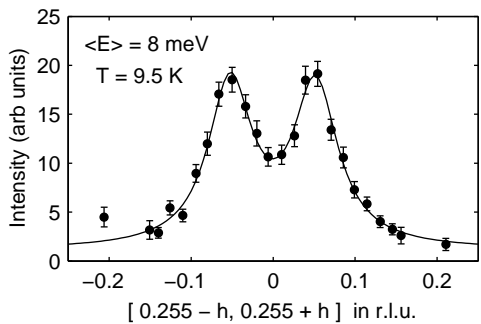}|
\caption{\small An example constant energy cut through the
magnetic zone centre $(0.255,0.255)$ in the ($-h,h$) direction:
$E$ = 8 meV. The solid curve is a fit to two Lorentzian peaks.}
\label{fig:cut}
\end{figure}

Figure \ref{fig:cut} shows a cut through unpolarized neutron data
in the ($h, -h$) direction. The data were measured at $T$ = 9.5 K
with an incident energy of 50 meV. The cut has been averaged over
$7\leq E\leq 9$ meV and $0.215\leq h\leq 0.295$ in the $(h, h)$
direction. Lorentzian profiles were fitted to the two peaks, which
correspond to spin-waves propagating parallel to the $(-h,h)$
direction away from the $(0.255, 0.255)$ magnetic zone centre.

\begin{figure}
\includegraphics*{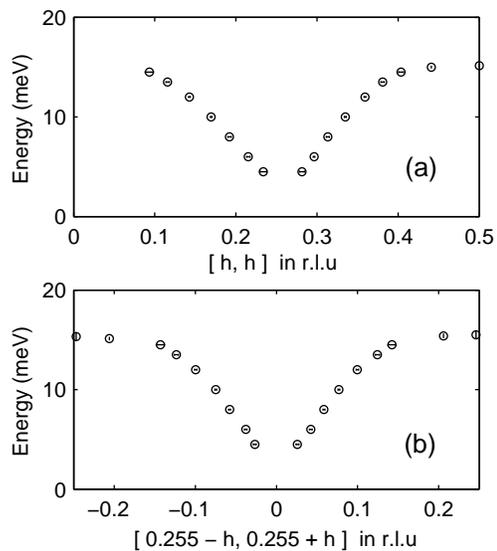}|
\caption{\small Spin-wave dispersion from the magnetic zone centre
at $(0.255, 0.255)$, measured at T = 9.5 K. (a) In the ($h,h$)
direction; (b) In the ($-h,h$) direction.} \label{fig:dispersion}
\end{figure}

By fitting (mainly) constant-energy cuts we determined the
two-dimensional spin-wave dispersions in the ($h,h$) and ($-h,h$)
directions. These are shown in figure \ref{fig:dispersion}. The
spin-wave velocity is found to be virtually isotropic in the {\it
ab} plane. The spin-wave  is seen to extend to $\sim$ 16 meV, in
contrast to $\sim$ 50 meV in La$_{1.5}$Sr$_{0.5}$NiO$_4$
\cite{freeman2}. This indicates a much weaker exchange interaction
in the cobaltates than in the nickelates. A quantitative analysis
of the spin-wave dispersion is in progress.

\begin{figure}
\includegraphics*{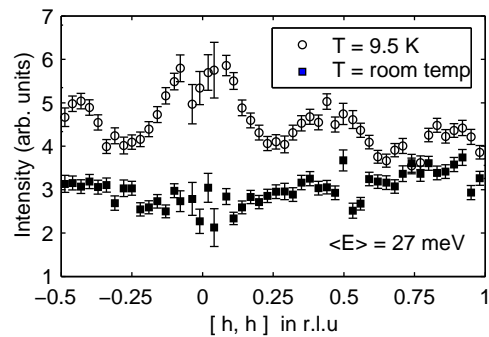}|
\caption{\small Diffuse magnetic scattering at T = 9.5 K, compared
with data at room temperature. Data was measured on MAPS and has
been averaged over $25\leq E\leq 29$ meV} \label{fig:diffuse}
\end{figure}

Another feature of the excitation spectrum is a pattern of diffuse
scattering between $\sim$ 25 meV and 33 meV, centred close to
positions such as $(0,0)$, $(\pm 0.5,0)$, $(0,\pm 0.5)$ and $(\pm
0.5,\pm 0.5)$. Figure \ref{fig:diffuse} depicts a cut along
$(h,h)$, with data averaged over $25\leq E\leq 29$. Scattering is
seen at 9.5 K but not at room temperature, which indicates that
the diffuse features are not due to phonons. Polarised neutron
measurements confirmed that the scattering is magnetic.

The origin of this diffuse magnetic inelastic scattering is not
clear at present. One possibility is that the Co$^{2+}$ spins have
strong planar anisotropy, and that the diffuse feature corresponds
to the excitation of the out-of-plane spin component. Another is
that we are observing magnetic correlations among the Co$^{3+}$
sites. These possibilities are currently being assessed by
modelling of the spin excitations.

We would like to thank the Engineering and Physical Sciences
Research Council of Great Britain for their help in funding this
work.

\end{document}